\documentclass[prl,twocolumn,preprintnumbers,amsmath,amssymb,floats]{revtex4}
\usepackage{tipa}

\usepackage{graphicx}

\begin{document}

\title{High-energy scale revival and giant kink in the dispersion of a cuprate superconductor}
\author{B. P. Xie$^1$, K. Yang$^1$, D. W. Shen$^1$, J. F. Zhao$^1$, H. W. Ou$^1$, J.
Wei$^1$, S. Y. Gu$^1$,
 M. Arita$^2$, S. Qiao$^2$, H. Namatame$^2$, M. Taniguchi$^{2}$,
 N. Kaneko$^3$, H. Eisaki$^3$, Z. Q. Yang$^{1}$, and
 D.L. Feng$^{1}$} \email{dlfeng@fudan.edu.cn}
 \affiliation{$^1$Department of Physics, Applied Surface Physics State Key Laboratory,
 and Shanghai Laboratory of Advanced Materials, Fudan University, Shanghai 200433, China}
 \affiliation{$^2$Hiroshima Synchrotron Radiation Center and Graduate School of Science, Hiroshima University, Hiroshima 739-8526, Japan.}
 \affiliation{$^3$AIST, 1-1-1 Central 2, Umezono, Tsukuba, Ibaraki, 305-8568 Japan.}
\date{\today}
\begin{abstract}
In the present photoemission study of a cuprate superconductor
Bi$_{1.74}$Pb$_{0.38}$Sr$_{1.88}$CuO$_{6+\delta}$, we discovered a
large scale dispersion of the lowest band, which unexpectedly
follows the band structure calculation very well. The incoherent
nature of the spectra suggests that the hopping-dominated dispersion
occurs possibly with the assistance of local spin correlations. A
giant kink in the dispersion is observed, and the complete
self-energy containing all interaction information is extracted for
a doped cuprate in the low energy region. These results recovered
significant missing pieces in our current understanding of the
electronic structure of cuprates.

\end{abstract}

\maketitle

The interplay between different energy scales of a physical system
usually holds the keys of its various fascinating properties. For
cuprate superconductors and their parent Mott-insulator, the
bandwidth of the so-called Zhang-Rice singlet band
\cite{ZhangRice} has been intriguingly illustrated in numerical
studies and
experiments\cite{Wells,FilipScience,DagottoPRB90,Tohyama,DahnkenPRB04}
to be $2.2J$ instead of $4t$, $J$ being the magnetic exchange
interaction, and $t$ being the hopping integral. This
renormalization of the effective bandwidth from $4t$ to $2.2J$ in
cuprates was considered to be an important step toward the
ultimate understanding of the high temperature
superconductivity\cite{Shenscale}, which highlights the importance
of magnetism. In practice, such effective band has been widely
adopted in various self-energy analysis of the low energy kinks in
the cuprate dispersion in terms of either
electron-phonon\cite{Alenature,Zhou,Cuk} or
electron-magnon\cite{JohnsonPRL,Valla05} interactions.

However, the above effective bandwidth largely depends on an
extrapolation in doped cuprate, since the quasiparticle dispersion
in a large region near the Brillouin zone center, $\Gamma$, has
not been identified so far, possibly due to very broad lineshape.
Based on an effective tight binding model \cite{Normantt}, hopping
parameters could be fitted out of the partial quasiparticle
dispersion observed in the vicinity of the Fermi surface up to
0.35\,eV below the Fermi energy $E_F$. The bandwidth of the
occupied part is then extrapolated to be about
0.4\,eV.\cite{Yangkeoverdope,Normantt}  Recent technological
advances enable us to study electronic structure in a much wider
momentum window. It is therefore pertinent to reexamine the
``missing dispersion" near $\Gamma$, and check whether this
extrapolation is justified. In this paper, we report the discovery
of such missing dispersion in the photoemission data of a cuprate
superconductor Bi$_{1.74}$Pb$_{0.38}$Sr$_{1.88}$CuO$_{6+\delta}$
(Bi2201). We found that sections of coherent and incoherent
spectra piece together a full dispersion actually on the order of
$4t$ \textit{again}. This large scale dispersion unexpectedly
follows the band structure calculation very well, except that
there is a giant kink at low energies. These findings provide a
fresh picture on the subtle interplay of different energy scales
in cuprates. Moreover, the experimental bare band was retrieved
from the data, which enables the experimental retraction of the
full self-energy of a doped cuprate. So far, only partial
self-energy has been studied in most previous experiments.
Therefore, the revelation of such a critical quantity that
contains the complete interaction information would provide really
strong constraints on theory.

\begin{figure*}[]
 \includegraphics[width=16cm]{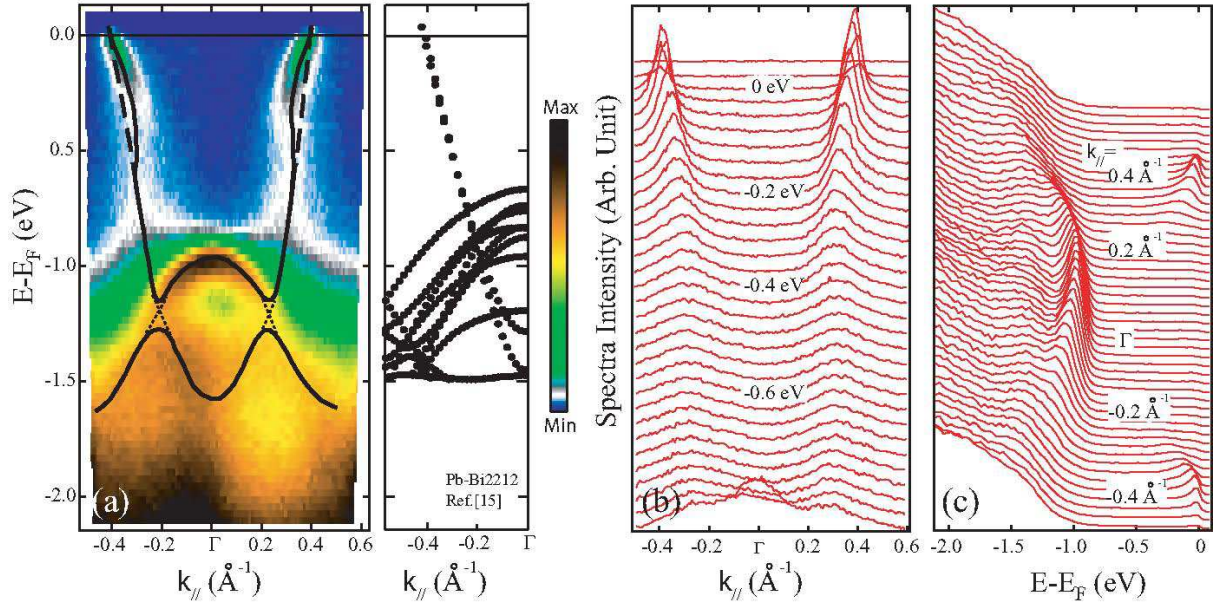}
\caption{Large energy scale dispersion of Bi2201. (a), Photoemission
intensity as a function of energy and momentum along the nodal
direction. The thick lines indicate the experimentally measured
dispersion. The thick dashed lines indicate the linear behavior of
the bare band as predicted by LDA calculations. The slight asymmetry
of the data are due to matrix element effects and possible small
misalignment from the nodal direction. The right side shows the LDA
band structure calculated up to 1.5\,eV for the nodal direction of
Bi2212 from Ref.[15]. (b), Selected momentum distribution curves,
and (c), Energy distribution curves of data plotted in panel (a).}
\label{f1}
\end{figure*}

Highly overdoped Bi$_{1.74}$Pb$_{0.38}$Sr$_{1.88}$CuO$_{6+\delta}$
single crystal was prepared by floating-zone technique and
subsequent annealing. The superconducting phase transition
temperature $T_c$ is 5K, indicating the hole concentration is very
close to the superconductor/metal phase boundary. ARPES
experiments were performed at the Beamline 9 of Hiroshima
Synchrotron Radiation Center (HiSOR), which is equipped with a
Scienta R4000 electron analyzer. The angular resolution is
$0.3^\circ$ and the energy resolution is 10\,meV. The data were
taken at $20$\,K with 22.5\,eV photons in the wide angle mode,
which covers more than $30^\circ$. This enables the identification
of broad features in momentum distribution curves, which is
crucial for the observation made here. The samples were aligned by
Laue diffraction, and cleaved in-situ in ultra-high vacuum better
than $5\times 10^{-11}\,mbar$.

ARPES intensity map for a highly overdoped Bi2201 in a wide energy
and momentum range is shown in Fig. 1a along the $(0,0)-(\pi,\pi)$
or nodal direction. Guided by the thick lines, one can clearly
observe two hybridized bands. The low energy one would have
dispersed down to $\sim1.6$\,eV below $E_F$ (\textit{i.e., }$4t$,
\textit{instead of }$2.2J$), if it had not intersected the high
energy band around $-1.2$\,eV. Because the broad MDC feature is
still much narrower than the wide angular window of the newly
developed analyzer, artifacts such as the inhomogeneity across the
two dimensional detector can be excluded. The dispersion within
[-0.9\,eV, -0.4\,eV] is clearly visible in the momentum distribution
curves (MDC's) (Fig.1b). The dispersion and smooth connection to the
strong feature near $\Gamma$ confirm that it is not caused by some
secondary electron background. Therefore, the strong feature around
$\Gamma$ is part of the same band as the quasi-particle band near
$E_F$. On the other hand, their relation is unobvious in the energy
distribution curves (EDC's) in Fig.1c, where the dispersion within
[-0.9\,eV, -0.4\,eV] cannot be resolved. The band and dispersion
energy scales identified in Fig. 1a agree with the recent local
density functional (LDA) band structure calculation of Pb-doped
bilayer Bi$_{2}$Sr$_{2}$CaCu$_2$O$_{8+\delta}$ (Bi2212) compound
very well (right side of Fig. 1a)\cite{BansilLDA}, as the bilayer
splitting is minimal along the nodal direction\cite{FengBBS}. The
strong feature near $\Gamma$ is shown mostly consisted of other Cu
$d$ and O $p$ orbitals than those for the band close to $E_F$. The
well preserved quasiparticles at such high energies (Fig. 1c)
indicate correlation effects are quite weak for these orbitals. The
interesting difference is that the experimental dispersion near
$E_F$ deviates from the straight line in the band calculation, and
exhibits a large kink.

This wide dispersion exists over a large region of the Brillouin
zone. As shown in Fig.2, besides the nodal cut in Fig. 2a, it is
also observed in cut \#2 and \#3, which is half way between the
nodal and antinodal regions. In cut \#4 and \#5, the renormalized
band is located near $E_F$. A weak feature also exists at high
energy, although there is no missing band dispersion. However,
compared with cut \#1-3, the high energy feature in cut \#4-5 is
weaker, and extends straight into high energies, does not show a
connection with the high energy band. On the other hand, the broad
lineshape and the resemblance do suggest the high energy features
observed in cut \#1-3 contain significant incoherent weight.
Recently, high energy dispersion beyond the $2.2J$ and incoherent
lineshape in the Mott insulator Ca$_2$CuO$_2$Cl$_2$ were reported
\cite{RonninghighE}, which have also been observed in various
numerical studies of hole motion in an antiferromagnetic spin
background\cite{DahnkenPRB04,WengPRB01,GroberPRB00}. The hole
concentration of the heavily overdoped system under study is about
1/4, so that long range antiferromagnetic order no longer exists,
and even long range spin fluctuations diminish\cite{Wakimoto}.
However, the neighboring sites of a hole mostly are still occupied
by spins, and the short range spin correlations are still strong,
even though the quasiparticle peak near $E_F$ is very well
defined\cite{Yangkeoverdope,Kaminskioverdope}. Consistently, the MDC
widths at high binding energies are around 0.2 \AA$^{-1}$, which
correspond a short length scale of 5 \AA. It has been shown that a
hole dressed with heavy local spin-flip cloud would enhance the
electron-phonon interaction greatly\cite{NagaosaPolaron}. In the
strong electron-phonon coupling regime, the spectrum could be shown
to be dominated by the incoherent part due to multiple phonon
excitations, and yet follows the bare band structure dispersion,
while the quasiparticles are renormalized to the vicinity of $E_F$
with little weight\cite{GunnarssonPolaron,KMO}. Lately, the broad
lineshape in the undoped La$_2$CuO$_4$ has been well simulated when
strong electron-phonon interactions were considered\cite{RoschPRL}.
Therefore, the observed bare band dispersion of the incoherent
spectral weight in doped cuprates could possibly be attributed to
strong electron-phonon interactions enhanced by local spin flip
cloud around the hole \cite{Yangkeoverdope,NagaosaPolaron}.  While
for the quasiparticles near -1\,eV, as these high energy orbitals
are less affected by the spin-flip cloud at the low energy orbitals,
the hole can travel more freely. Consequently, an anomalous
electronic structure for Bi2201 is revealed: well defined coherent
quasiparticles exist at both high energies and low energies, giving
a total occupied band width of 1.2\,eV, while incoherent features
that follow the bare band dispersion dominate the intermediate
energy region. This anomalous dispersion covers most of the
Brillouin zone as shown by the dark region in the inset of Fig.2.

\begin{figure}[t]
 \includegraphics[width=8.5cm]{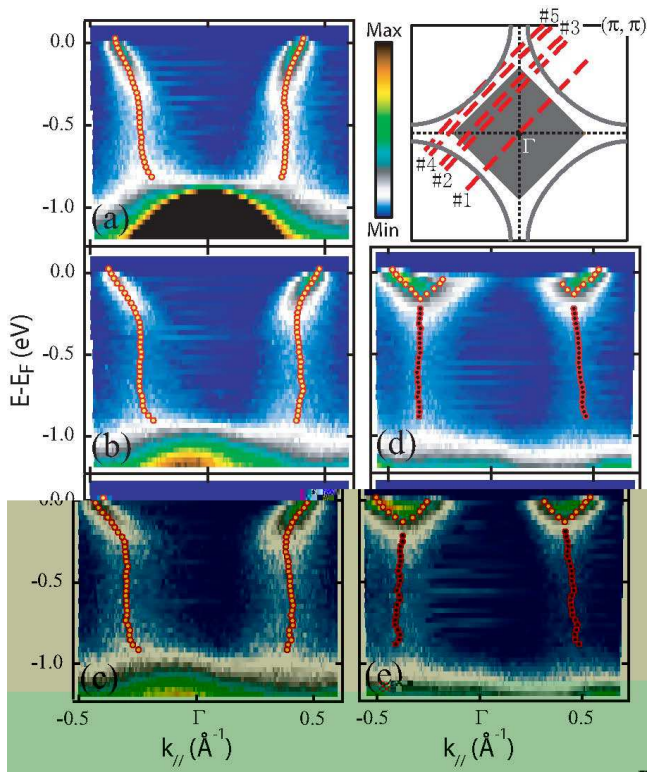}
 \caption{Momentum
dependence of the high energy dispersion. (a-e) Photoemission
intensity taken along the cut \#1-5 respectively in the Brillouin
zone as illustrated in the inset on the top right corner, where
large energy scale dispersion happens in the dark region. The yellow
dots mark the position of the measured band dispersion, while the
black dots in (d) and (e) indicate the centroid of the incoherent
feature.} \label{f2}
\end{figure}

\begin{figure}[t]
 \includegraphics[width=8.5cm]{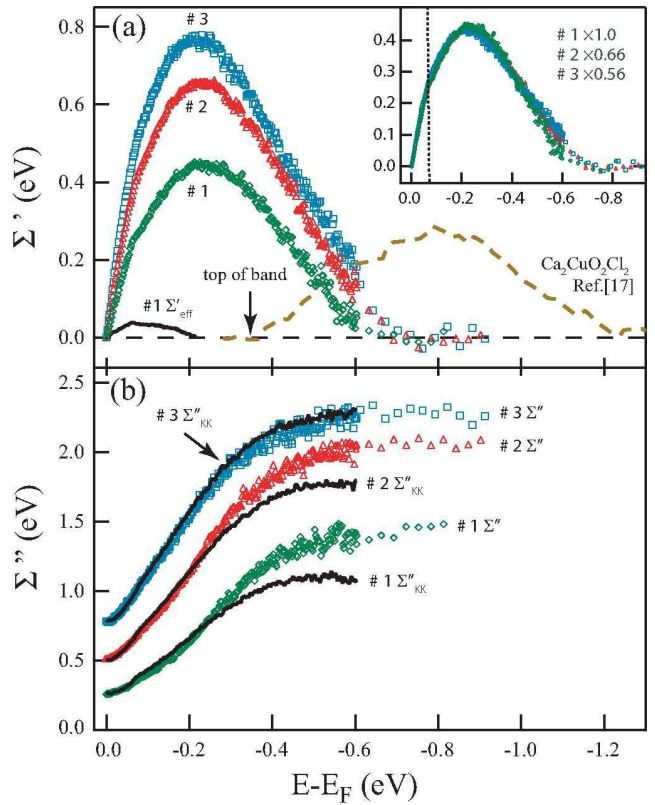}
\caption{Full self-energy of the Bi2201 extracted from the measured
dispersion. (a) the real part of the self energy for cut \#1
(diamonds), \#2 (triangles), and \#3 (squares). The solid line is
real part of the self energy for the conventional low energy kink,
$\Sigma_{eff}^{'}$. The dashed line is the high energy kink observed
in parent Mott insulator compound Ca$_2$CuO$_2$Cl$_2$ for comparison
purpose\cite{RonninghighE}, the arrow marks the top of its band.
Inset: The $\Sigma^{'}$ for cut \#1-3 could be re-scaled to match
each other, where the dashed line indicates the low energy kink
position. (b) Imaginary part of the full self-energy as a function
binding energy. The data were shifted up by 0.25\,eV and 0.5\,eV for
cut\#2 and \#3 respectively. The solid curves are the imaginary part
calculated from the real part of the self-energy based on
Kramers-Kronig transformation.} \label{f3}
\end{figure}

One remarkable observation in Figs. 1 and 2 is the giant kink
structure in the dispersion compared to the band structure
calculation. The large scale dispersion observed here and the linear
 dispersion behavior near $E_F$ predicted by LDA calculations provide a rare
chance to extract the full self-energy, $\Sigma$, of a strongly
correlated system, which reflect all correlation effects in the
system. Fig. 3a shows the real part of the self-energy, $\Sigma^{'}$
which is the difference between the bare band (\textit{e.g.} the
dashed line in Fig. 1a) and the measured band. $\Sigma^{'}$ contains
contributions from all interactions. In previous studies, because of
the lack of bare band information, a fraction of the self energy,
$\Sigma_{eff}^{'}$ was retrieved by assuming a local effective band
structure\cite{ZhouReview}. The resulting partial self-energy was
argued to contain information on interactions between electrons and
certain bosons. As shown in Fig. 3a, $\Sigma_{eff}^{'}$ is only a
small fraction of $\Sigma^{'}$. Away from the nodal cut,
$\Sigma^{'}$ increases toward the antinodal direction.
Interestingly, as shown in the inset of Fig. 3a, $\Sigma^{'}$ of
different cuts can be scaled to a universal curve, where even the
details of the low energy kink structures matches each other. These
raise the questions whether the low energy kink is just a part of a
large scale electronic effects, and whether the kink extracted in
conventional analysis cleanly represents the interaction between
electrons and bosons\cite{Alenature,Zhou,Cuk,JohnsonPRL,Valla05}.
The $\Sigma^{'}$ extracted from the high energy dispersion data of
Ca$_2$CuO$_2$Cl$_2$ is also plotted in Fig.3a for
comparison\cite{RonninghighE}. The qualitative resemblance between a
Mott insulator and a highly overdoped cuprate here remarkably
illustrates their similar short range behavior. On the other hand,
the observed doping dependency also provides additional evidences
for further refinement of the role that spins and lattice play in
the large energy scale dispersion.

The imaginary part of the self-energy $\Sigma^{''}$ is shown in Fig.
3b, which is the product of the MDC width and the bare band velocity
(the slope of the dashed line in Fig. 1a). Generally, it increases
rapidly in the first 0.35\,eV, then gradually saturates at high
binding energies, indicating that the spectrum is dominated by
incoherent excitations here. For comparison, $\Sigma^{''}_{KK}$, the
Kramers-Kronig transformation of $\Sigma^{'}$, is shown
here.\cite{FinkKK} The good agreement between $\Sigma^{''}$ and
$\Sigma^{''}_{KK}$ on the low energy side for all three cuts
confirms that the full self-energy has been reliably extracted at
least in the first 300\,meV below $E_F$. On the other hand, the
deviation at high energies again reflects the incoherent nature of
the electronic structure there. We note that the large amplitude and
energy expansion of $\Sigma$ at such high doping is quite anomalous.
The self energy measured here provide direct and critical
information for the development of theory for cuprates.


To summarize, we have presented a global picture of the electronic
structure and the interplay of $t$ and $J$ in a highly overdoped
cuprate. Contrary to previous perceptions, the band width is still
determined by $t$, although the low energy quasiparticles exist in
the $J$ energy scale. On the other hand, the high energy dispersion
over a large fraction of the Brillouin zone may still be attributed
to $J$ through the polaronic effects induced by strong local spin
correlations. Moreover, various intriguing new information is
revealed under this global picture, such as the much larger
amplitude and energy expansion of the giant kink compared with the
low energy kink discussed before, the universal scaling of
$\Sigma^{'}$, and resemblance to the insulator. Since the high
energy and local physics set the footing for low energy properties,
these findings provide a comprehensive view of the different energy
scales in the problem. More over, the full self-energy extracted for
the the low energy ($<$300 meV)  put strong quantitative constraints
on models that are designed to understand high temperature
superconductivity in a realistic cuprate material.

D.L.F. would like to thank Profs. Z. Y. Weng, and J. P. Hu for
helpful discussions. This work was supported by NSFC and MOST (973
project No:2006CB601002) of China, Science \& Technology committee
of Shanghai, and the Core-University Program of JSPS.

\end{document}